\documentstyle[12pt,epsfig]{article}

\def\etal{{\it et al.}}
\def\etc{{\it etc.}}
\def\ie{{\it i.e.}}

\begin{document}
\setcounter{footnote}{0}
\renewcommand{\thefootnote}{\fnsymbol{footnote}}

\begin{flushright}
SLAC--PUB--8323 \\
December 1999
\end{flushright}

\bigskip\bigskip\bigskip
\begin{center}
{\Large Assessment and Outlook}\footnote{Work supported by the Department
of Energy under contract number DE-AC03-76SF00515.}\\[3ex]
{\large Burton Richter} \\[1ex]
{\it Stanford Linear Accelerator Center} \\
{\it Stanford, California  94309}
\end{center}
\thispagestyle{empty}

\vfill
\centerline{Concluding Remarks presented at the}
\centerline{XIX International
Symposium on Lepton and Photon Interactions at High Energies}
\centerline{Stanford University, August 9-14, 1999}
\vfill\eject

\section{Introduction}

My assignment at this conference is to assess where we are in
high-energy physics and speculate on where we might be going.  This
frees me from any obligation to summarize all that went on here and
allows me to talk just about those topics that interest me the most at
this time.  I will let my prejudices show and talk a bit about physics
in general, CP violation, neutrinos, accelerators, non-accelerator
experiments, and even theory.

I start with a bit of my own history in physics and what I see as cycles
in science---I think these are relevant.  I began research in physics
when I was in my third under-graduate year at MIT. I worked with
Professor Francis Bitter in his magnet laboratory and learned how to do
an experiment.  I spent three years with Francis Bitter, but, as I began
my second year of graduate school, I found myself becoming less
interested in the nuclear magnetic moments that I was trying to measure
and more interested in the fundamentals of the protons and neutrons that
contributed to the moments.  I shifted to particle physics, never
regretting it.

When in graduate school I did both theory and experiment.  It was not
hard to do that back then because, if you could manipulate $\gamma$
matrices, set up and solve integral equations and understood what a
Green's function was, you could keep up with theory and even do some of
your own.  I think that an experimenter in graduate school today would
not have an easy time keeping up with the sophisticated mathematics
required to understand what's going on in string theory (but, I think it
would be worth a try).

In the 1950s, virtually every major university had an accelerator of
its own, including MIT, Harvard, Yale, Columbia, Cornell, Princeton,
Chicago, the University of Illinois, Michigan, Minnesota, Berkeley, Caltech,
 \etc\   The really big facilities were the 6-GeV Bevatron at
Berkeley, and the 3-GeV Cosmotron at the Brookhaven National Laboratory.
I learned about accelerators at a time when the experimenters and the
accelerator builders were the same people.  When I was a graduate
student, four of us maintained, modified and ran the MIT synchrotron.
Accelerators have become much more sophisticated since then and our
field has become more specialized.  The accelerator builders have split
off from the experimenters just as the experimenters split off from the
theorists a hundred years ago.

Particle physics as practiced has changed enormously.  Small experiments
of two or three people that were the norm when I started out have given
way to huge collaborations culminating in the ATLAS and CMS
collaborations at the LHC, which have 1500 to 2000 collaborators each.
These collaborations are larger than many of our laboratories.

Accelerators are huge and are built by specialists.  There are very few
of them.  The theory has become so mathematically sophisticated that it
is very difficult to keep up.  An experimental physicist at the height
of his or her powers (that means as a graduate student or early post
doc) would have a difficult time with topology, knots, branes, and all
of today's machinery.  However, the questions really remain the same as
they were when I was a graduate student:  what are the fundamental
entities, what determines their properties, what governs their
interactions, how did it begin and how is it going to end?

Our science, like all sciences, repeatedly goes through a three-part
cycle.  In one phase, experiment leads theory.  The discoveries come
thick and fast, and no model exists to absorb and interpret them.  That
was the situation in the 1950s when meson and baryon resonances were
proliferating and it became clear that these could not be the
fundamental entities from which all matter was built because there were
too many of them.

The second and perhaps most enjoyable phase is the time when experiment
and theory advance rapidly together.  The 1960s and 1970s were like
that, when the experimenters and theorists played a kind of ping-pong
game when an experiment would lead to a new theoretical interpretation
which led to a new experimental test which led to a new modification of
the theory.  In quick order over a period of ten years, the standard
model rose from deep inelastic scattering, scaling, quarks, neutral
currents,  $\psi/J$, the third generation, the GIM mechanism, renormalizable
gauge theories, \etc

Now we are in a stressful and frustrating phase.  We have an obviously
incomplete model and no experiment definitively points the way toward
the next step.  The standard model stands uncontradicted, yet we know
perfectly well that it is wrong:  it has too many arbitrary constants,
18 of them if you do not include the seven more that may come from
neutrino masses and a neutrino-mixing matrix.  It is interesting to note
that there are more arbitrary constants in the standard model now than
there were resonances in the particle data book when it became the
accepted view that there were too many resonances for them to be
fundamental.  There are other problems as well:  there is not enough CP
violation in the standard model to create the baryon asymmetry of the
universe and there are potential problems with longitudinal $W$
scattering.

We are at a point in high-energy physics where we have a wonderful model
with its quarks, leptons, force particles, and its $SU(3) \times SU(2) \times U(1)$
structure.  All that is currently accessible to experiment has been
correctly predicted by the theory (except for the 18 to 25 constants, of
course).  Yet there are things that the standard model cannot deal with
and, at all our conferences in recent years, we have hoped to hear the
experiments and theoretical ideas that will lead us beyond it.

\section{CP Violation}

Many of the advances in high-energy physics, indeed in all of science,
come from the overthrow of unexamined assumptions.  Such is the history
of CP violation.  The first assumption to go was that of parity
conservation, a theoretical simplification which had been experimentally
shown to be correct for the electromagnetic and nuclear forces.  It was
then assumed that it was correct for the weak interaction as well.

In the early 1950s, high-energy physics was struggling with the $\theta$-$\tau$
paradox.  Experiments had turned up what seemed to be two spin-zero
particles, one of which decayed into two $\pi$-mesons while the other
decayed into three $\pi$-mesons.  Both appeared to have the same mass to a
precision of a few tenths of a percent.  Lee and Yang, in their famous
paper \cite{ref:a}, analyzed what we really knew about
parity conservation in the weak interactions and concluded that there
were no constraints.  They gave several examples of experimental tests
that could settle the question, and C. S. Wu did one of those, studying
nuclear beta decay and finding that parity was not conserved.  Very soon
thereafter experiments at the Columbia and Chicago cyclotrons found that
parity was also not conserved in the $\pi$-$\mu$-electron decay chain.  Fitch
and Cronin in their experiment of 1966 found that CP was not conserved
in kaon decay, and ever since we have been trying to understand what is
going on.

So far, all we know about CP violation comes from the study of the
$K$-meson system.  That should change next year when the asymmetric
$B$-Factories at SLAC and KEK produce enough data for significant
measurement of CP violation in the $B$-meson system.  Both machines and
detectors have started up very well and, at the time of this writing
(November 1999), PEP-II has reached a luminosity of $1.4 \times10^{33}
{\rm cm}^{-2} s^{-1}$,
about 40\%\ of design.  KEK-B is not far behind.  It is reasonable to
expect that roughly 10 $fb^{-1}$ of data will be accumulated by the time of
the year 2000 Rochester Conference in Kyoto, which should be enough to
achieve an error of about $\pm 0.15$ in sin$2\beta$ from the $\psi$-$K_s$ channel.

There are new results from studies of the $K$-meson system that address
the existence of CP violation outside the CKM matrix.  Both CERN (NA-48)
and FNAL (K-TeV) groups have new results measuring $\epsilon'/\epsilon$.
These results are given together with their previous results in Table\  1.  The weighted
average of the results has a $\chi^2$ per degree of freedom of 2.8, which
gives a confidence level of less than 3\%, an improbable result.  The
particle data group in such cases scales the errors to get the
appropriate $\chi^2$ per degree of freedom, a procedure that some may object
to.  I do it anyway and, with the error scaled, $\epsilon'/\epsilon$ is still not
consistent with superweak interactions.
\begin{table}[htb]
\begin{center}
\begin{tabular}{|l|c|c|}
\hline
Experiment & $\epsilon'/\epsilon\ (10^{-4})$ & Comment  \\
\hline\hline
E-731      & 7.4 $\pm$ 5.9   & FNAL 1993 \\ \hline
NA-31    & 23.0 $\pm$ 6.5 & CERN 1993 \\ \hline
K-TeV   & 28.0 $\pm$ 4.1  & FNAL 1999 \\ \hline
NA-48    & 18.5 $\pm$ 7.3  & CERN 1999 \\ \hline
Average  & 21.2 $\pm$ 2.8  & $\chi^2/{\rm dof} = 2.8$ \\ \hline
Average  & 21.2 $\pm$ 4.7  & Scaled Errors \\ \hline
\end{tabular}
\caption{High-precision $\epsilon'/\epsilon$ Measurements}
\label{tab:1}
\end{center}
\end{table}

The low confidence level of these results illustrates a problem that we
all should think carefully about; are we handling errors properly?  The
two FNAL experiments, for example, have a probability of only one in 300
of coming from a random set with gaussianly distributed errors.  The
Fermilab group is led by the most careful physicist I know, Bruce
Winstein, and the low probability may simply be bad luck.  There is,
however, another possibility and that is that we don't really fully
understand the error distribution functions.

All of us in high-energy physics are guilty of treating a collection of
systematic errors as if they were random gaussianly distributed errors,
a procedure that we know is wrong.  However, we don't know how to do it
any better.  We also know that even if an individual error is random and
gaussianly distributed, ratios of sums and differences of such
quantities may not be gaussianly distributed.  There are also
non-gaussian tails on acceptance functions, tracking functions, \etc\   We
need to understand this problem better but, until we do, complex
experiments with many variables, complicated triggers, and many cuts in
the analysis process, should perhaps be treated with some caution.
Treating errors in these complicated experiments as if they were
gaussian may lead us to ascribe a much higher confidence level to a
conclusion than is really deserved.

\section{Neutrinos}

Certainly one of the most exciting areas of research at present is
neutrino physics.  It is fair to say that the results of the last decade
on the neutrinos from the sun, from the atmospheric interaction of
cosmic rays, and from accelerators, are changing our thinking and
challenging the standard model.  There are new data from Super
Kamiokande (Super-K) on solar neutrinos; new data from Super-K and
others on atmospheric neutrinos; and still a problem with the Los Alamos
experiment (LSND) which doesn't seem to fit well with our current
prejudices.

Before discussing the present situation, I want to mention two people
who are not currently involved in the program, but who played an
absolutely critical role in the evolution of neutrino physics.  The
first of these is Dr.  Raymond Davis, Jr., now retired, of Brookhaven
National Laboratory.  Ray Davis had an idea that at first seemed
impossible, to detect neutrinos from the sun as a way to find out about
the nuclear physics of the solar cycle.  As far as I can tell, the
original idea of using the chlorine-argon inverse beta-decay reaction to
detect neutrinos goes back to a paper by Bruno Pontecorvo \cite{ref:b}
written in 1946.   In that paper, Pontecorvo discussed ways to
detect neutrinos, including neutrinos from the sun.  He dismissed the
solar idea because the flux would be too small for a one-cubic-meter
detector which was the largest that he could think of.

Ray Davis thought much bigger.  In 1955, he began working at Brookhaven
with a 1000-gallon chlorine detector.  He found no events because of the
small size of the detector but, with this apparatus, he perfected the
argon-chlorine radio-chemical separation techniques that allowed his
later experiments to succeed.  In the early 1960s, he began
construction of a 100,000-gallon detector at the Homestake Mine in South
Dakota and began collecting data in 1967.  From the very beginning his
results indicated that the flux of neutrinos from the sun was less than
that predicted by the solar models.  This was greeted at first with
considerable skepticism, but over the years Davis' results began to be
taken more seriously and, with the results of the SAGE and GALLEX
experiments, the reality of the neutrino deficit has been fully
accepted.  Davis' work opened this field.

The second person I want to mention is Professor Masatoshi Koshiba, now
retired from the University of Tokyo.  In 1979, Koshiba proposed the
construction of the huge water Cerenkov counter that became known as the
Kamiokande Detector.  Construction was completed in 1983 and initial
experiments focused on the search for proton decay.  By 1985 the
simplest $SU(5)$ version of grand unified theories had been ruled out.
With the addition of an outer veto layer to the detector beginning in
1984, Kamiokande became capable of detecting neutrinos as well.
Kamiokande measured the solar neutrino flux, confirming Davis' result,
and also detected neutrinos from supernova SN1997A, opening up a new
field of neutrino astrophysics.  Even before much in the way of results
from Kamiokande had come in, Koshiba proposed, in 1983, the construction
of the very much larger Super-K.  Under his leadership the project was
approved just before his retirement from the University of Tokyo.  The
results from Super-K are what are generating all the excitement about
atmospheric neutrinos.

\begin{table}[htb]
\begin{center}
\begin{tabular}{|l|c|c|}
\hline
Experiment & Threshold Energy & Ratio to Solar Model \\
\hline\hline
Homestake Mind    & 817 keV    & 0.33 $\pm$ 0.03 \\ \hline
SAGE                    & 235 keV   & 0.52 $\pm$ 0.07 \\ \hline
GALLEX               & 235 keV   & 0.59 $\pm$ 0.06 \\ \hline
Kamiokande           & 7 MeV    & 0.54 $\pm$ 0.07 \\ \hline
Super-Kamiokande & 5.5 MeV  & 0.48 $\pm$ 0.02 \\ \hline
\end{tabular}
\caption{Results of Solar Neutrino Experiments}
\label{tab:2}
\end{center}
\end{table}

The results of the solar neutrino experiments are summarized in Table 2.
The SAGE and GALLEX experiments with their 235-kV neutrino energy
threshold are the only ones that are sensitive to the proton-proton part
of the solar cycle where almost all of the energy of the sun is
produced.  The Homestake experiment (Ray Davis') with its 800-kV
threshold is sensitive to the beryllium-7 line plus the boron-8
continuum.  The Kamioka and Super-K experiments are sensitive only to
the boron-8 continuum.  The surprising, to me, result is the Homestake
experiment which gives a ratio to the solar model of one-third, while
the other four give a ratio of one-half.  Any explanation of these
results is complicated by the required energy dependence of the neutrino
depletion process needed to account for all of the results.  Super-K
also reports a statistically marginal distortion of the neutrino
spectrum compared to the standard solar model, and a 1-1/2 to 2 standard
deviation day-night effect.  Neither of these two effects is as yet of
the significance required to be taken seriously.

\begin{figure}[htbp]
\begin{center}
{\epsfbox{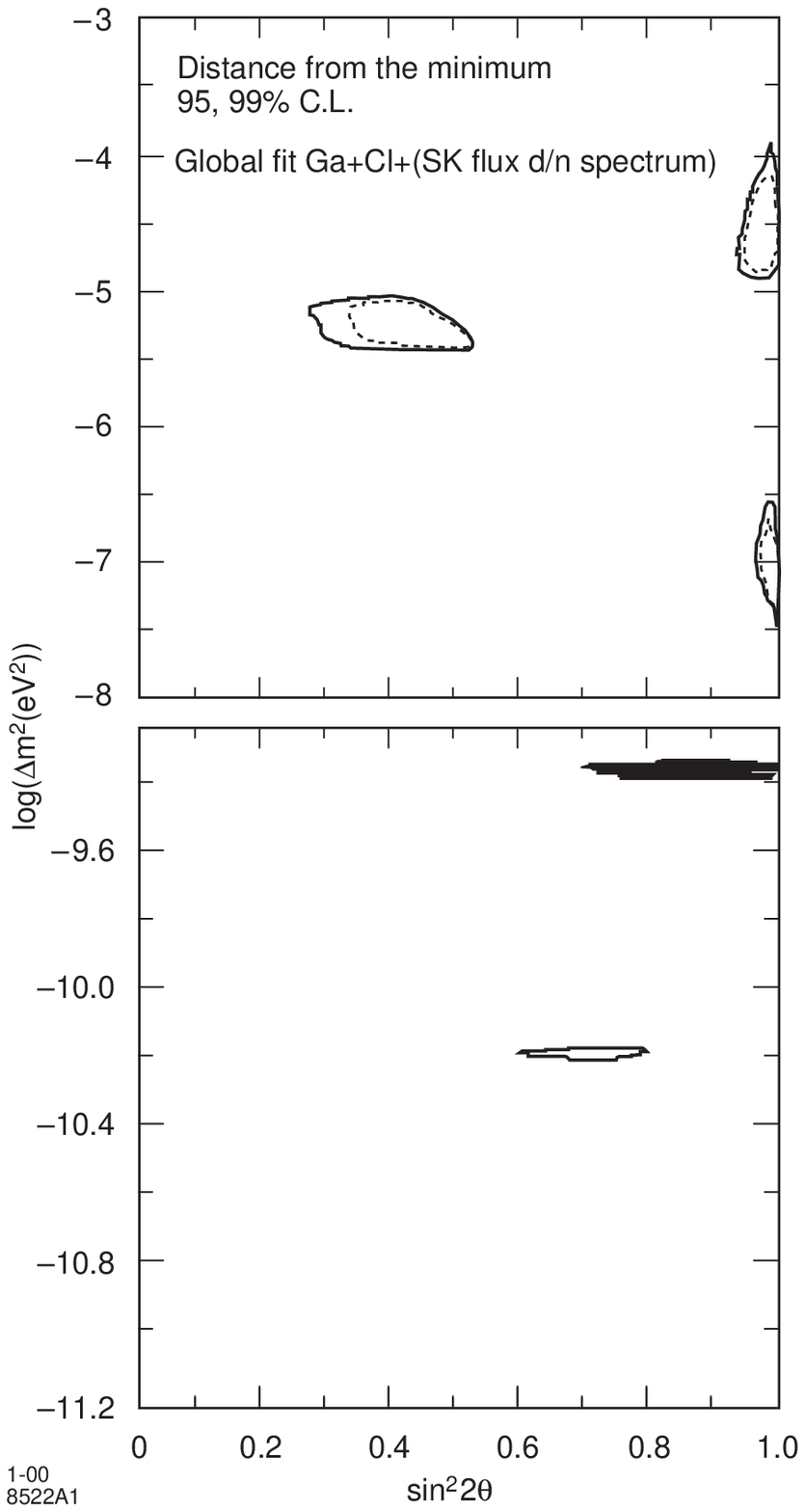}}
\caption{Global fit to the solar neutrino data  \cite{ref:c}. }
\label{fig:1}
\end{center}
\end{figure}

Figure 1 is from Y. Suzuki's talk \cite{ref:c} at this conference summarizing the
solar neutrino situation within the framework of neutrino oscillations.
The upper part of the figure gives the allowed regions in the MSW model
(a resonant conversion of the electron-neutrinos into another species of
neutrinos) while the lower figure gives the allowed regions for pure
vacuum oscillations.  It is worth noting that if there is a systematic
effect in the Homestake experiment, and all of the results were to be
consistent with an energy-independent reduction of the solar neutrino
flux, the large mixing-angle solution would be allowed with any $\Delta m^2$
from about $10^{-9}$ to $10^{-4}\ {\rm eV}^2$.

The LSND experiment has been a problem since the first results were
presented in 1995.  It is the only experiment that purports to show the
conversion of muon neutrinos into electron neutrinos with a large $\Delta m^2$.
The first results appeared to be in conflict with other experiments,
particularly the KARMEN experiment at the Rutherford Laboratory.  There
is more data now, and as DiLella \cite{ref:d} showed in his talk, there appears now
to be a narrow range where the results of all of these experiments are
consistent.  Figure 2  shows the present situation.

\begin{figure}[htb]
\begin{center}
{\epsfxsize=3.5in\epsfbox{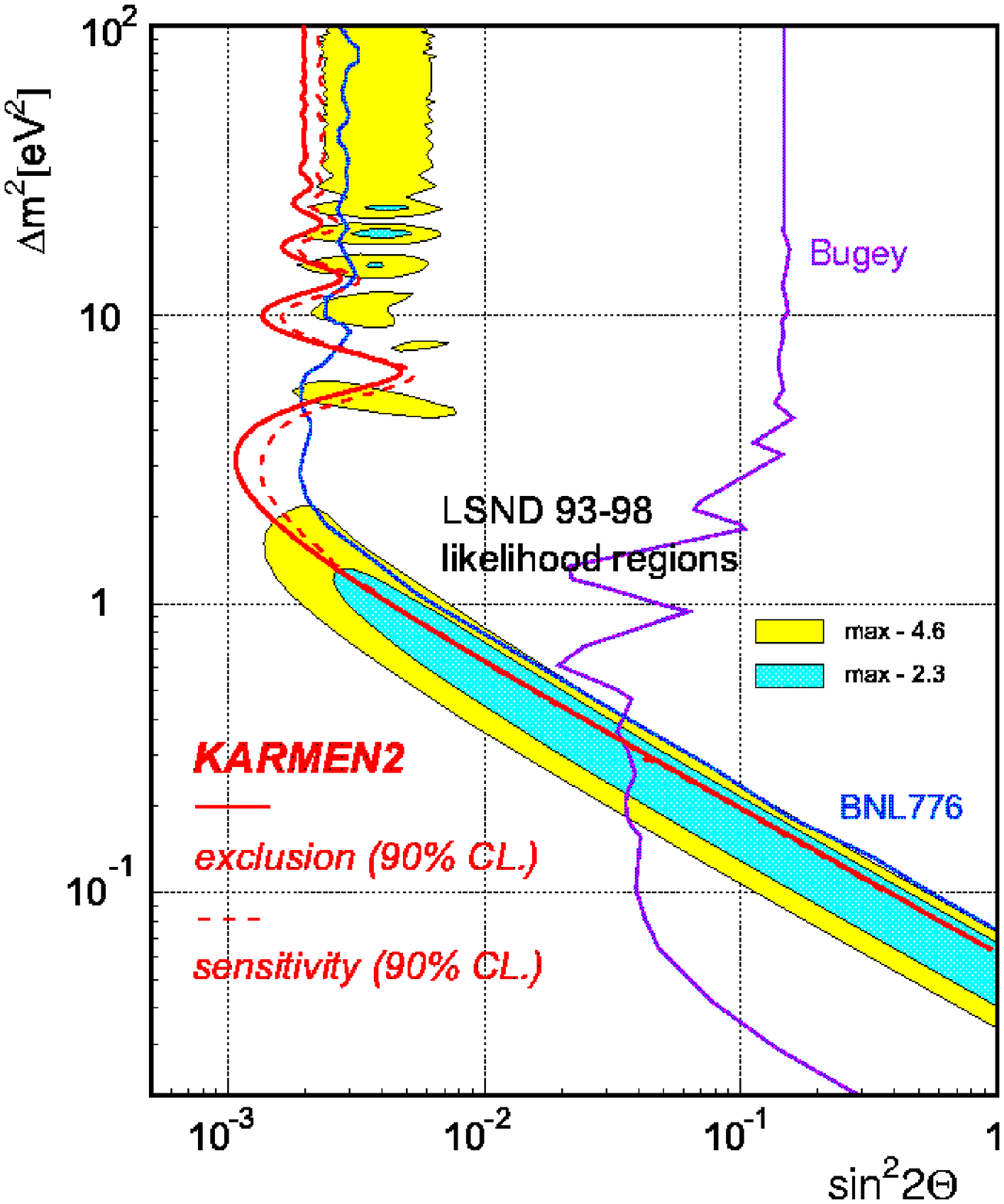}}
\caption{Electron neutrino experimental results.  LSND is not in conflict with
other experiments in a band from $\Delta m^2$ of 0.3 eV$^2$ and sin$^22\theta$
of $3 \times 10^{-2}$ to $\Delta m^2$ of l eV$^2$ and sin$^22\theta$ of $1.5 
\times 10^{-3}$  \cite{ref:d}.}
\label{fig:2}
\end{center}
\end{figure}

The LSND experiment uses muon antineutrinos arising from the decay of
mesons produced at the LAMPF 1-GeV proton accelerator.  If these muon
antineutrinos oscillate into electron antineutrinos, these can interact
with protons in the detector tank (behind a thick shield) to produce
positrons (20  to 60 MeV cuts) plus a neutron which is detected after a
delay by a neutron-captured gamma ray.  Electron neutrinos, which can be
produced directly in the beam, can also interact in the detector, but
they interact with carbon and there is no delayed neutron-captured gamma
ray.  The experiment analyzing all the data from 1993 through 1998,
claims to see an excess of positrons in coincidence with a delayed
neutron capture of 40 $\pm$ 9 events.  That is certainly statistically
significant, even with the statistical skepticism I evidenced earlier,
but there are still many doubters and I include myself among the
doubters.

\begin{figure}[htb]
\begin{center}
{\epsfxsize=5in\epsfbox{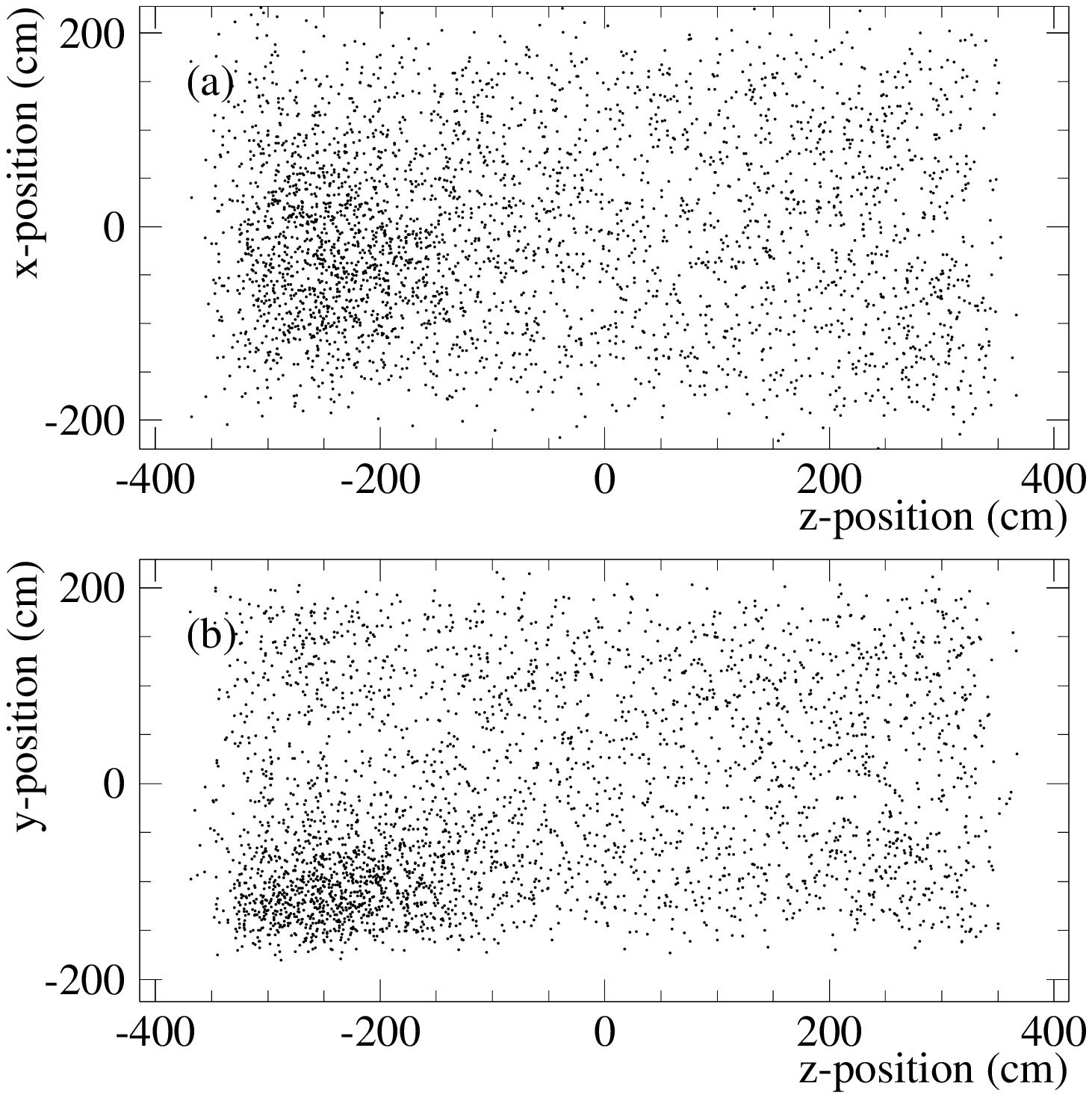}}
\caption{Spatial distribution of accidental background events in the LSND
detector. }
\label{fig:3}
\end{center}
\end{figure}

Figure 3 is from the LSND paper \cite{ref:e} and shows the distribution of accidental
events in their detector tank.  There is clearly a large excess of such
events at the bottom front of the tank and this can only come from
neutron leakage under their shield.  This is a ``beam on" background that
the experimenters eliminate by focusing on a fiducial region toward the
back and above the bottom of the detector.

Neutron diffusion is much less of a potential problem in the KARMAN
experiment, because their machine produces a 10 $\mu s$ beam pulse while
LAMPF produces a 600  $\mu s$ pulse.  Neutron diffusion under the shield is a
slow process and so KARMAN can gate most of this out with a relatively
narrow time window around the beam pulse.

\begin{figure}[htb]
\begin{center}
{\epsfxsize=3in\epsfbox{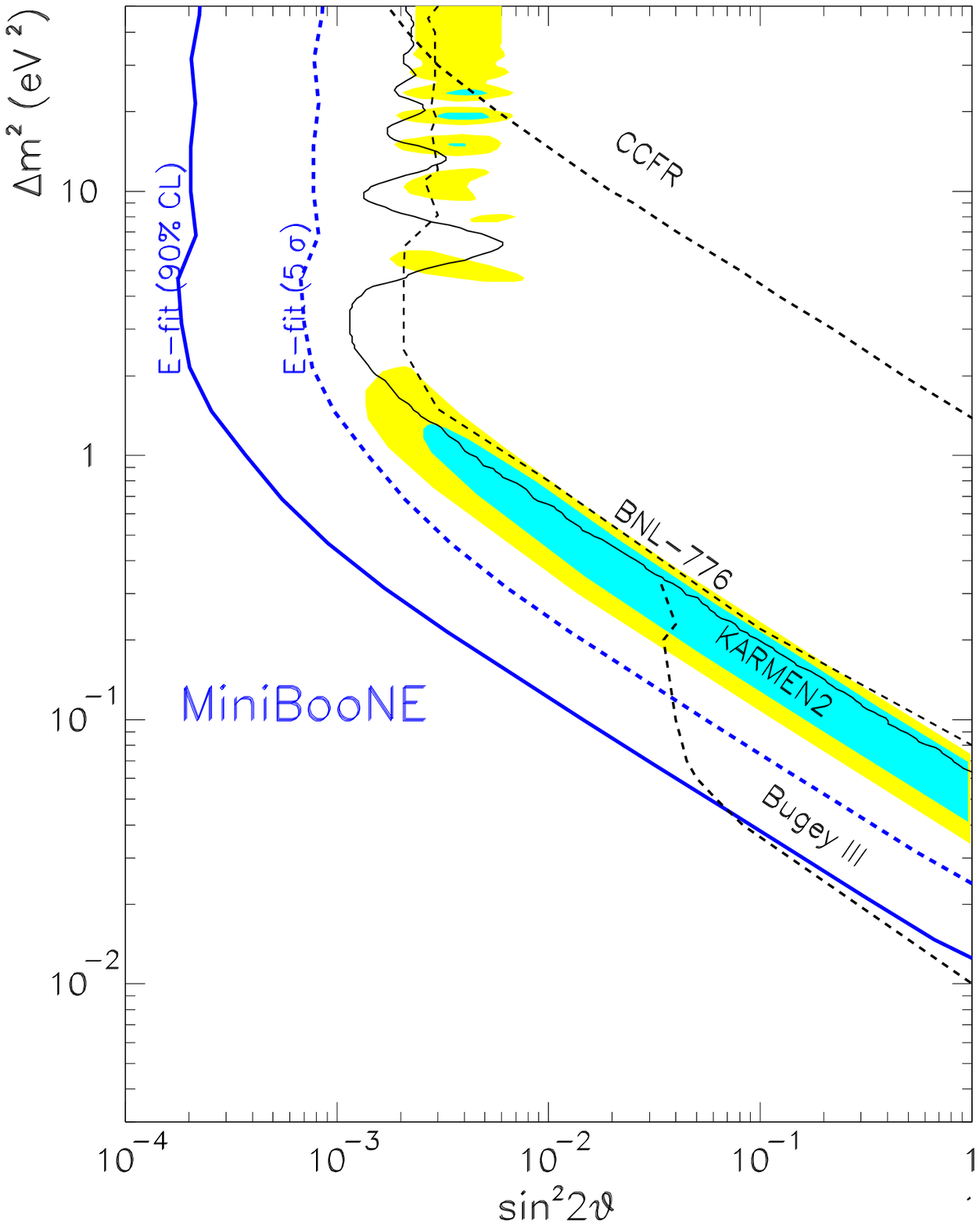}}
\caption{Expected range of the MiniBooNE experiment in the electron neutrino 
$\Delta m^2$ versus sin$^22\theta$ plain superimposed on the present results.  
Data taking is scheduled to begin in 2002  \cite{ref:d}.}
\label{fig:4}
\end{center}
\end{figure}

This experiment will have to be done again and it will be done again at
FNAL in the MiniBooNE.  The MiniBooNE detection limits are shown in
Fig. 4 and the first results should be available in the year 2003.  We
have to wait.

\begin{figure}[htb]
\begin{center}
{\epsfxsize=4.5in\epsfbox{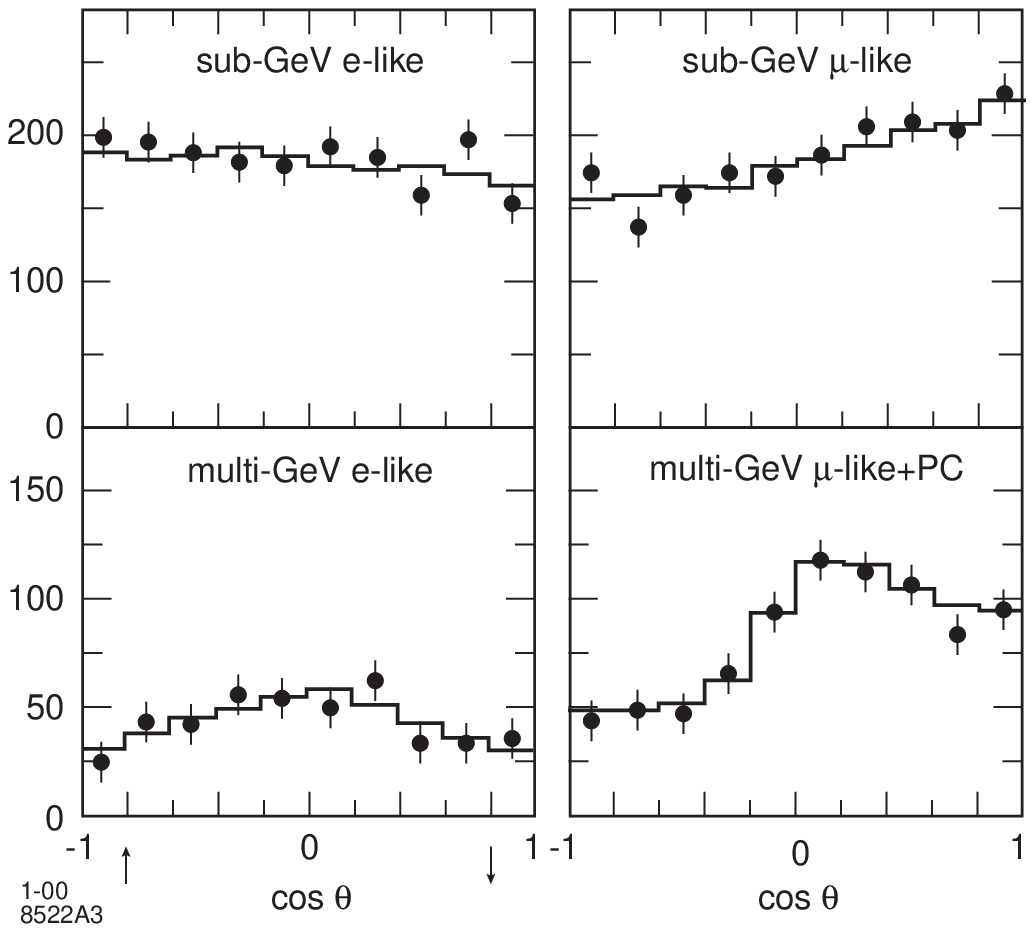}}
\caption{Zenith angle distribution of the Super Kamiokande data for electron-like
and muon-like neutrino events.  The muon-neutrino-like events are
strongly depleted for upward-going neutrinos  \cite{ref:f}.}
\label{fig:5}
\end{center}
\end{figure}

The most exciting news on the atmospheric neutrino front comes from the
Super-K data.  Figure 5 \cite{ref:f}  shows an apparent disappearance
of muon neutrinos as a function of zenith angle.  Muon-like events are
strongly depleted when they are generated by neutrinos passing through
the entire earth compared to those generated by neutrinos coming down through the
relatively thin covering of the Super-K detector.  The allowed mass
difference region assuming $\nu_\mu$ to $\nu_\tau$ oscillations is shown in Fig. 6.
The Macro and Soudan experiments see a similar effect though with much
looser mass constraints.  Note that the Super-K/Macro/Soudan data alone
do not tell us what happens to the muon neutrinos, they only tell us
that they disappear.  Muon neutrinos could go to sterile neutrinos that
don't interact; $\tau$ neutrinos that can only generate neutral current
events since all the neutrinos are below $\tau$ production threshold; or
even to some extent into electron neutrinos since it is hard to tell an
enhancement of electron neutrinos from a shortage of muon neutrinos in
this data.

\begin{figure}[htb]
\begin{center}
{\epsfxsize=3.5in\epsfbox{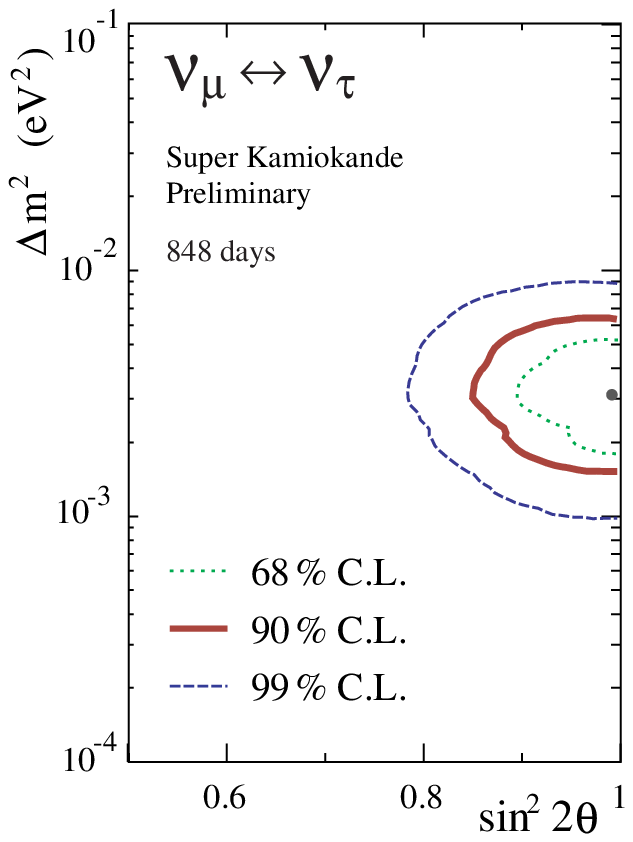}}
\caption{Allowed $\Delta m^2$ and sin$^22\theta$ from the Super Kamiokande data
assuming $\mu$-$\tau$ neutrino oscillations \cite{ref:f}.}
\label{fig:6}
\end{center}
\end{figure}

\begin{figure}[htbp]
\begin{center}
{\epsfxsize=4in\epsfysize=6in\epsfbox{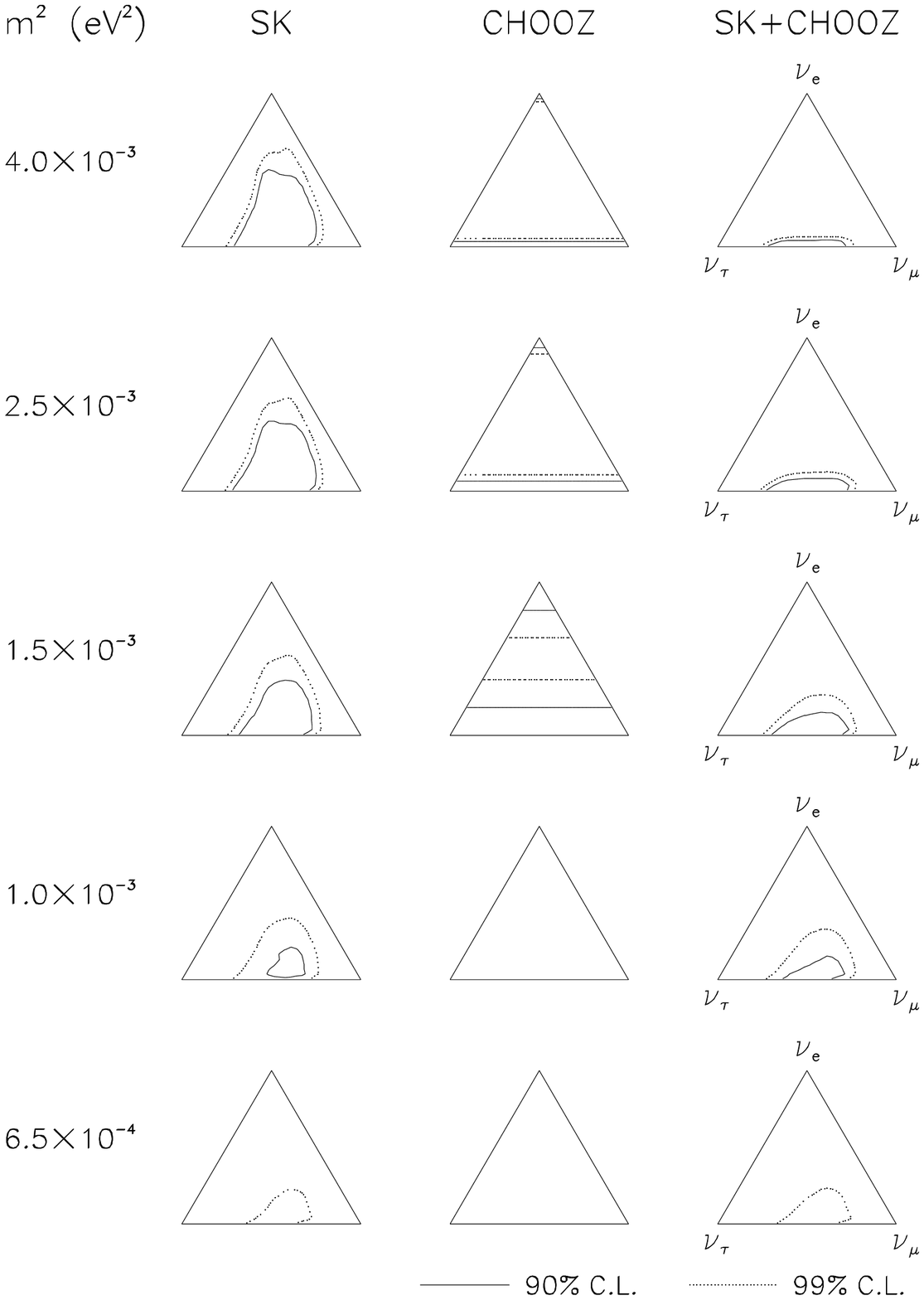}}
\caption[]{The Fogli, \etal, analysis showing how the CHOOZ electron-neutrino
disappearance limit together with the Super Kamiokande results limit the
electron-neutrino/muon-neutrino coupling \cite{ref:g}. }
\label{fig:7}
\end{center}
\end{figure}

The CHOOZ reactor experiment, in conjunction with the Super-K data,
imposes the tightest restrictions on muon neutrino electron neutrino
mixing.  Fogli \etal\  \cite{ref:g}
analyzes Super-K and CHOOZ data together (Fig. 7).  The shaded regions
give the 90\%\ and 99\%\ allowed regions for the two experiments.  The
analysis done is done in the ``dominant mass" mode and Super-K alone
allows a quite large mixing of electron neutrinos.  Super-K combined
with CHOOZ, however, says that if there are oscillations, the muon
neutrino oscillates almost entirely into tau neutrinos.  Another reactor
experiment, the Palo Verde experiment, has only been running for a
relatively short time but already confirms the CHOOZ result in the mass
range favored by Super-K.

Before discussing the next generation of experiments, it is useful to
pause to summarize what we know:

\begin{enumerate}
\item
Neutrinos from the sun are below the predictions of the standard solar
model.  All of the experiments integrate over neutrino energies above
some threshold.  The SAGE and GALLEX experiments, with a threshold of
235 keV, are a factor of two below the expectations; the Homestake
experiment, with a threshold of 817 keV, is a factor of three below
expectation; and the Kamiokande and Super-K experiments, with thresholds
of 5 to 7 MeV, are a factor of two below expectations.

\item
The atmospheric
neutrino experiments clearly see an azimuthal dependence of the
muon-neutrino/electron-neutrino ratio indicating a decrease of muon
neutrinos coming 12,000 km through the earth.

\item
The LSND experiment
claims to see muon-neutrino to electron-neutrino conversion, but this
experiment needs confirmation.

\item
The CHOOZ experiment is the most
sensitive of the reactor experiments and sees no loss of
reactor-generated electron neutrinos over distances in the order of
kilometers.

\end{enumerate}

\noindent
Taking all of this together, the favored explanation is neutrino-flavor
oscillations plus the MSW effect.  However, this is not the only
explanation consistent with the data we have so far.  For example,
Barger \etal\  \cite{ref:h} proposed to
explain the data with a mixture of neutrino decay and the MSW effect.
Gonzales-Garcia \etal\  \cite{ref:i} hypothesized
flavor changing in neutral current interactions of neutrinos.  Both of these
alternative hypotheses invoke new physics but so does the favored explanation.
The job of the next generation of experiments is to sort all of this out.

On the solar neutrino front, we will certainly get more data from
Super-K and data from two new experiments, SNO (already operating) and
Borexino (due to start up in one to two years).  I doubt that more
statistics from Super-K will tell us anything new, but the other two
experiments certainly will.  Of particular interest to me is Borexino,
which should be able to resolve the Be$^7$ line and so pin down any energy
dependence with some precision.

The KamLAND reactor experiment in Japan is also important.  Its
electron neutrinos come from twelve nuclear reactors at an average
distance of 150\  km.  KamLAND is a conversion of the old Kamiokande
experiment and it should have a sensitivity to electron-neutrino
oscillations down to $\Delta m^2$ less than
$10^{-5} \ {\rm eV}^2$, covering the large
mixing-angle MSW solution to the solar neutrino deficit.

Super-K will get much more data on atmospheric neutrinos.  Of particular
interest would be data around 90 degrees from the vertical.  If
oscillation is the answer and the mass difference is around Super-K's
central value, their sub-GeV and multi-GeV samples will show different
behavior in this region.  The 0.5-1 GeV neutrinos would be near their
first oscillation minimum at 200-km distance, while the above 1-GeV
neutrinos would show little reduction.  Super-K may not have enough
flux, consistent with the required angular resolution, to do this
analysis.

The K2K experiment takes muon-neutrinos from KEK's 12 GeV proton
synchrotron to the Super Kamiokande detector and has just begun taking
data.  This is a disappearance experiment using muon neutrinos of less
than 2 GeV.  If the mass difference is above about 2 $\times 10^{-3}\ {\rm eV}^2$ they
should clearly see an effect in two to three years of data taking.  This
would be the first independent check of the Super-K result.  In
principle it is possible for the K2K experiment to definitively confirm
the oscillation hypothesis.  At the central value of the Super-K mass
difference, 500~MeV neutrinos oscillate away and 250 MeV neutrinos
oscillate back.  However, the neutrino flux and the Super-K energy
resolution are probably not good enough to see this, but it would be
wonderful if they were.

I have already mentioned the MiniBooNE experiment.  It supplies a
definitive check on LSND and we will have to wait a few years for
results.

The long baseline experiment, MINOS, is under construction at Fermilab
and at the Soudan Mine.  It is expected to begin data taking in the year
2003.  MINOS is capable of seeing tau neutrino appearances over most of
the mass difference range allowed by Super-K.  Under MINOS' conditions,
neutrinos of 2 GeV will oscillate away and  1\ GeV will return.  If I were
running the experiment, I would certainly tune the beam to low-energy
neutrinos for the first few years.

CERN and the Gran Sasso Laboratory in Italy are discussing a European
long-baseline experiment.  The beam and the distance are very similar to
the MINOS experiment.  If this experiment is approved, I hope the
apparatus is sufficiently different from MINOS to make the investment
worthwhile.

\begin{figure}[htb]
\begin{center}
{\epsfxsize=3.5in\epsfbox{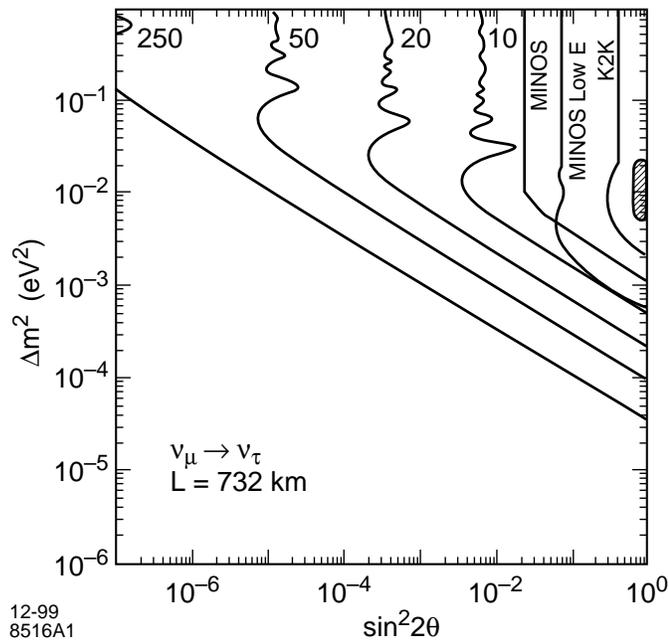}}
\caption{The Barger, \etal\ \cite{ref:h}, calculation of the limits that could be
placed on $\nu_\tau$
appearance using a muon storage-ring neutrino source of various muon
energies.}
\label{fig:8}
\end{center}
\end{figure}

Finally, there is much discussion of the potential of a muon storage
ring as a high-intensity neutrino source.  Such a facility would be much
simpler to build than the muon collider that has been discussed for the
past few years, but there are issues that couple the storage-ring design
to the experiment.  These issues have not yet been fully explored.
Figure 8 \cite{ref:j} shows what
the excitement is all about; lots of neutrinos and greatly improved
sensitivity.  Figure 9 shows the problems; the storage ring is a mixed
source of, for example, muon-neutrinos and electron-antineutrons.

\begin{figure}[htbp]
\begin{center}
{\epsfxsize=3.5in\epsfbox{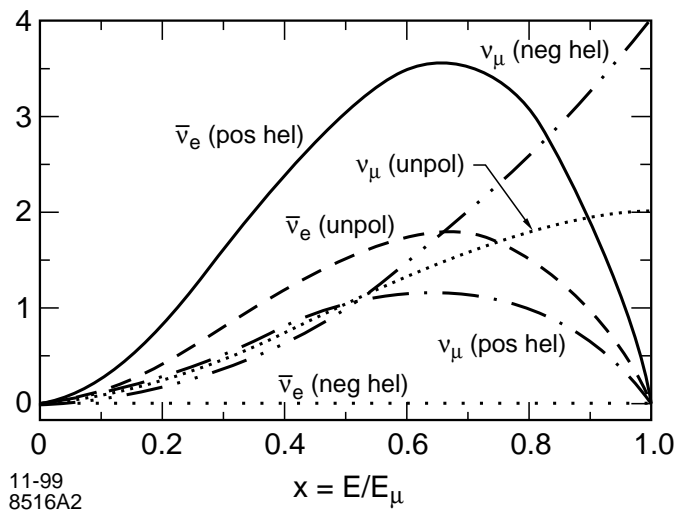}}
\caption{Normalized neutrino flux versus fractional neutrino energy at zero
degrees for various negative muon helicity states  \cite{ref:k}.}
\label{fig:9}
\end{center}
\end{figure}

The muon and electron neutrino spectra from unpolarized muons in the
storage ring (much easier for the machine builder) are not that
different, and only a detector with really good energy resolution could
separate them on a statistical basis.  To complicate the situation
further, if the muon beam in the storage ring has a transverse momentum
comparable to the muon mass (this gives the highest flux), then the two
spectra are smeared further making the two types still harder to
separate.

Polarization of muons in the storage ring allows the electron neutrinos
to be tuned all the way to zero (in principle).  However, the best
polarization that the machine designers have come up with so far is in
the range of 20-30\%\ and even achieving this costs considerably in
complexity and somewhat in flux in the machine.

Groups in the U.S. and in Europe are working on the machine and on the
physics.  They need to work more closely together to better understand
the trade-offs between the experiment and the machine so that both can
be properly optimized in order to carry out the required physics
program.

\section{Accelerators}

At this conference, J. Lykken \cite{ref:l} has spoken of the physics potential of the
next generation of proton and electron accelerators, and G. A. Voss \cite{ref:m} has
spoken of the state of technologies.  I can, therefore, be brief.  The
next big machine, the LHC, is under construction at CERN and is close to
being the first world project.  Contributions to the accelerator and
detector have been made by many nations outside the group of CERN member
states.  The LHC will have an energy of 14 TeV in the proton-proton
center of mass, a luminosity of 10$^{34} {\rm  cm}^{-2}s^{-1}$ and a mass
reach of about 1\  TeV.  Operations are expected to begin in the year 2005.

The two main experiments ATLAS and CMS will each have 1500 to 2000
collaborators.  The size of these collaborations is unprecedented and
presents difficult organizational problems in getting ready and new
sociological problems in operation.  In the 500-strong collaborations of
today, we already have a bureaucratic overlay to the science with
committees that decide on the trigger, data analysis procedures, error
analysis, speakers, paper publications, \etc\   The participating
scientists are imprisoned by golden bars of consensus.  While we have
survived so far and preserved some opportunity for scientific
initiative, this will become more difficult as the collaborations grow
to three times the size of today's largest.  This needs thinking about
and talking about, but is a topic for another time.

The theorists tell us what they want us to find with the new
accelerators, the source of electroweak symmetry breaking.  But, that is
not the experimenters' job.  Our job is to find out what is really there
and, in this case, that means to find if there is any electroweak
symmetry making.  The experiments are difficult and the detectors are
complex, expensive devices.  Data rates, particularly at proton
colliders, are enormous and there is no way to digest it all.  Complex,
multi-tiered trigger systems are needed to reduce the flood of data
coming from the machine by a factor of 10 million or more so that our
computer systems can handle the load.  Those events that do not pass the
trigger screen are discarded.

There is a danger here.  Will we set up the experiments that can only
find what we expect to find?  Some years ago I asked the spokesperson
for a large experiment if they could see a magnetic monopole that decays
into a large number of 0.5 to 1\  GeV gamma rays (a model that was
plausible at the time).  The answer was ``no"; the trigger screened out
such isotropic events that had no jet-like clustering and no transverse
energy unbalance.  While this particular problem is fixed, there is a
lesson here.  Be prepared and allow some of the trigger to be dedicated
to what may currently be unfashionable.

\begin{figure}[htb]
\begin{center}
{\epsfxsize=3.5in\epsfbox{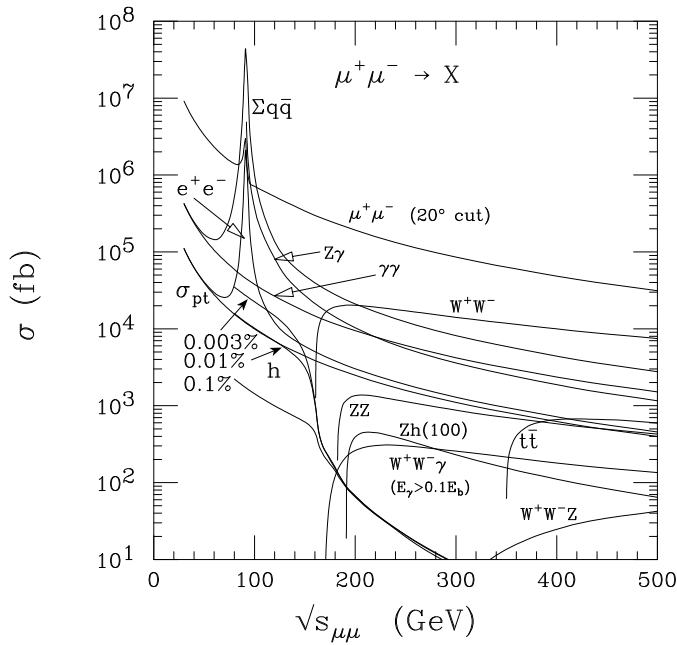}}
\caption{Lepton/anti-lepton cross-sections versus center-of-mass energy.  All
cross-sections except for $s$-channel Higgs production are the same for
$e^+$-$e^-$ and $\mu^+$-$\mu^-$  colliders \cite{than}.}
\label{fig:10}
\end{center}
\end{figure}

With the LHC well underway, the next accelerator facility will be an
electron-positron linear collider.  Lykken, in his presentation, told of
the physics potential and how the linear collider seems a necessary
compliment to the LHC.  Electron-positron machines have one advantage
over proton colliders:  all of the cross-sections are within a few
orders of magnitude of each other (Fig. 10) and it is still possible
to record all of the events that occur.  Beams in linear colliders can
be polarized and this is a great help in untangling the physics.
Research and development is nearly done and it is hoped that
construction of a new facility can start in 2004 or 2005 and the physics
program can begin in 2010.  The machine will be costly and will surely
have to be built by worldwide collaboration.

A critical issue not yet settled is the initial energy for a new
facility and what, if any, expansion capabilities should be built into
the initial design (spending some extra funds now can save time and
costs later for an energy increase).  There seems to be a convergence on
500\  GeV as an appropriate initial center-of-mass energy with the
potential to expand to something in the 1 to 1.5\  TeV region.  There is a
worldwide physics study currently underway with coordinators in Asia,
Europe and North America.  Whatever the initial energy, this project
will be expensive and will only happen if the entire high-energy physics
community gets behind it.

There has been much talk about the possibility of muon colliders.  The
attraction of the muon collider is also its biggest problem.
Synchrotron radiation, which may limit the performance of multi-TeV
electron-positron linear colliders, is absent in a muon collider.  This
absence requires the development of a new method of emittance damping to
shrink the phase space in which the muons are produced so that high
luminosity can be obtained in the collision.  This will require a major
R\&D effort and one is being planned.  It will not be simple.

The latest story on the state of the R\&D in systems design is given in
Ankenbrandt \etal\ \cite{ref:n}.   The
luminosity of such a collider appears not to be interesting until a
multi-TeV energy is reached.  The indicated low-energy option has
luminosity less than that of LEP-II, while the medium-energy example has
luminosity less than the linear collider.  All of the cross sections for
muon colliders are the same as those for an electron-positron collider
except for an s-channel process coupled to mass, Higgs-boson production.
Here also the muon collider is not as effective as the electron
collider.  The rate for the production of 100-GeV mass Higgs in the muon
collider is given in the reference above to be about 4000 per year.  The
500-GeV electron-positron machine produces five times that rate through
the Higgs plus $Z^0$ channel and the $W$-fusion channel.

The muon source itself is expensive.  Current studies use a 4-MW proton
source to produce the muons, four times the power of the SNS spallation
neutron source which is estimated to cost \$1.3 billion.  Perhaps this
can be done for less by the upgrade of one of the existing proton
machines, but any real test of the muon collider concept will require a
physics program that justifies the large cost.  Perhaps the muon storage
rings being discussed as neutrino sources can supply the justification
for a real trial of the technology.

\section{Non-Accelerator Experiments}

Non-accelerator experiments are playing an ever more important role in
high-energy physics; witness the time spent at this conference on Super
Kamiokande neutrinos, cosmic microwave background fluctuations, and
supernova distributions.  These experiments allow tests of our theories
not possible with accelerators.  More such experiments are coming, such
as:

\begin{itemize}
\item
The Sloan Digital Sky Survey which will map the large-scale structure of
luminous matter in the universe to much larger $z$ than now.  This
information is critical to test cold versus warm versus hot dark-matter
scenarios.

\item
The AUGER ultra-high energy cosmic ray experiment which will
get considerably more information on cosmic rays of mysterious origin
with energies beyond the cutoff from interaction with the $3^\circ K$ microwave
background radiation.

\item
The Whipple, Hegra, Egret, GLAST experiments,
\etc, looking for an explanation of the origin of ultra-high-energy
cosmic gamma rays.

\item
Neutrino observatories, under the ice and under the
oceans.

\item
Follow-ons to COBE and the Supernova Cosmology Project that
will give clues to the basic structure of the universe.

\end{itemize}

\noindent
Some in high-energy physics are concerned that such experiments may
drain funds from accelerator-based activities that have dominated
high-energy physics for many decades.  That may be so, but high-energy
physicists should go where the high-energy physics is, whether it be in
space, underground, underwater, or with accelerators.  There is more to
high-energy physics than the Higgs boson, and collaborations with the
astrophysicists and cosmologists are more likely to broaden rather than
narrow support for the things we are interested in through
collaborations with new (to us) areas of science.

Two particularly important programs have been discussed at this
conference, the cosmic-microwave-background radiation and the supernova
search.  The microwave- background radiation experiments measure the
fluctuations in temperature (on the order of $10^{-5\circ} K$) as a function of
the angular scale of the fluctuations.  The pattern of these temperature
fluctuations is a critical test of inflation, which ten years ago was
thought to be an untestable metaphysical concept.  It is testable now
(see, for example, C. Contaldi,  \etal\  \cite{ref:o}).

\begin{figure}[htb]
\begin{center}
{\epsfxsize=4in\epsfbox{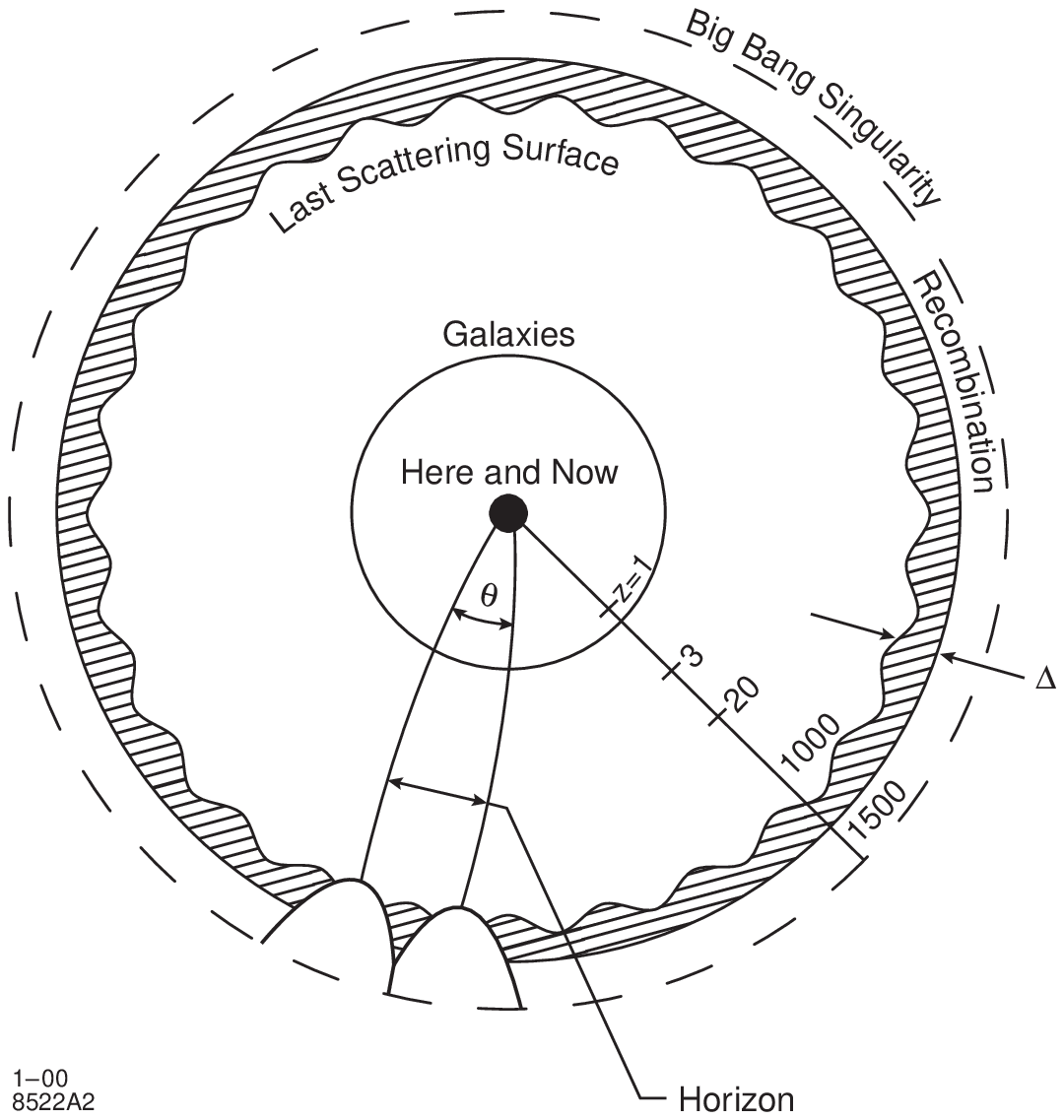}}
\caption{Schematic of the source of the fluctuations in temperature of the cosmic
microwave background radiation \cite{cobe}.}
\label{fig:11}
\end{center}
\end{figure}

The experiment looks back to about 300,000 years after the big bang when
the temperature in the early universe had dropped to the point where
electrons could form hydrogen atoms (see Fig. 11).  At that time the
mean-free path of light changed from very short to very long as free
electrons were bound to atoms; metaphorically the universe changed from
foggy to clear.  That light has been red- shifted by the Hubble
expansion of the universe to become the $3^\circ K$-background
microwave-background radiation.  Cold spots cooled early and hot spots
later.  The expansion leads to the fluctuations in temperature in the
microwave background.

\begin{figure}[htb]
\begin{center}
{\epsfxsize=3.5in\epsfbox{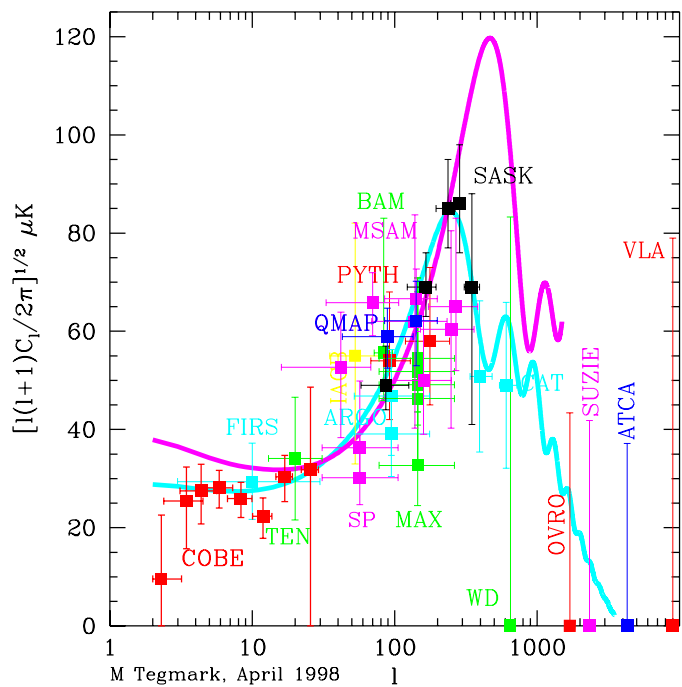}}
\caption{From Turner and Tyson  \cite{ref:p}.
Summary of current measurements of the power spectrum of CMB
temperature variations across the sky against spherical harmonic number
$\ell$ for several experiments.  The first acoustic peak is evident.
The lower curve, which is preferred by the data, is a flat Universe
($\Omega_0 = 1,\ \Omega_M = 0.35$), and the dark curve is for an
open Universe ($\Omega_0 = 0.3$) [courtesy of M. Tegmark].}
\label{fig:12}
\end{center}
\end{figure}

The available data are summarized in Fig. 12, which plots the power
spectrum of the fluctuations versus spherical harmonic number.  The
information needed for test of cosmology occurs at high-harmonic numbers
and the data is not precise enough yet to constrain the models.  Over
the next five years two new experiments will be launched, the MAP
satellite by NASA, and the Planck satellite by the European Space
Agency.  These experiments are designed to be remarkably precise.
The expected errors at large $\ell$ are expected to be comparable to the
width of the lines in Fig. 12 up to $\ell \approx 1000$ for MAP and 2000
for Planck.  Beyond these, the errors blow up very rapidly.
 It requires extraordinary precision to distinguish between
cosmological models and, when both satellites are up, I'll believe the
data from two independent satellites with two independent instruments.

Perlmutter \cite{ref:q} discussed the supernova cosmology
project which addresses one of the most fundamental questions in
physics, the constancy of the expansion rate of the universe.  The
results seem to say that it has not been constant, that the matter
density of the universe is less than the critical density expected, and
that there seems to be a cosmological constant in general relativity.

The experiment uses Type 1a supernovas as standard candles, measuring
the apparent brightness (distance by $R^{-2}$) and red shift $z$ (distance by
Hubble expansion).  In an unscientific sample of 15 high-energy
physicists, I found only one who understood how Type 1a supernovas
worked and what generates the light, and so I will briefly digress to
tell you about it.

Type 1a supernovas are thought to come from white dwarf stars in binary
systems that, over time, accrete matter from their companions until they
reach a critical mass (1.4 solar masses).  At that point, the white
dwarf collapses and explodes as a supernova.  Neutrinos escape
immediately (SN1987A, for example).  The debris cloud is heated by the
decay of radioactive nickel (the minimum of the nuclear- binding energy
curve), but the light is trapped because the cloud is optically opaque.
As the cloud expands, its optical thickness decreases ($R^{-2}$ for a uniform
sphere, and $R^{-1}$ for a shell), and the light intensity rises.  It falls
with the decay of the radioactive heat source.  Type 1a supernovas are
rare, occurring about one per 500 years per galaxy.  The collaboration
has collected fifty of them out to a \ $z$ of 0.8.

\begin{figure}[htbp]
\begin{center}
{\epsfxsize=3.5in\epsfbox{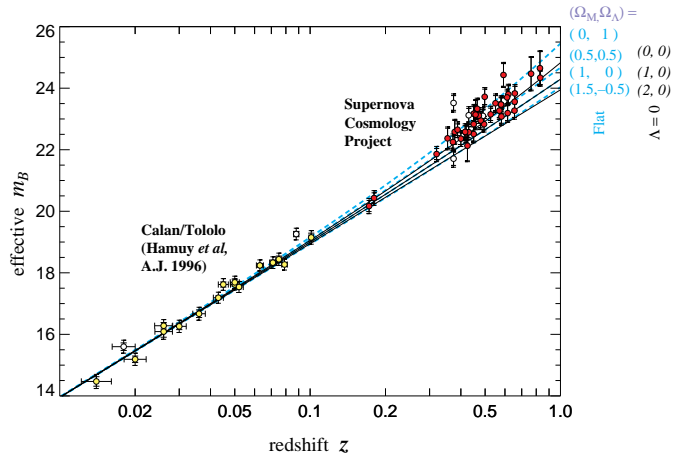}}
\caption{Effective magnitude versus redshift for Type 1a supernova.}
\label{fig:13}
\end{center}
\end{figure}

The data are shown in Fig. 13 that plots the apparent magnitude versus
red shift.  The data clearly deviate from the straight line expected for
a flat universe with a matter density equal to the critical density.
Perlmutter, \etal\ \cite{ref:q} find that for a flat universe the matter density is
0.28 $\pm$ 0.085 (statistical) $\pm$ 0.05 (systematic) of that expected for a
flat universe.  The rest has to be made up by a cosmological constant.

\begin{figure}[htb]
\begin{center}
{\epsfxsize=3in\epsfbox{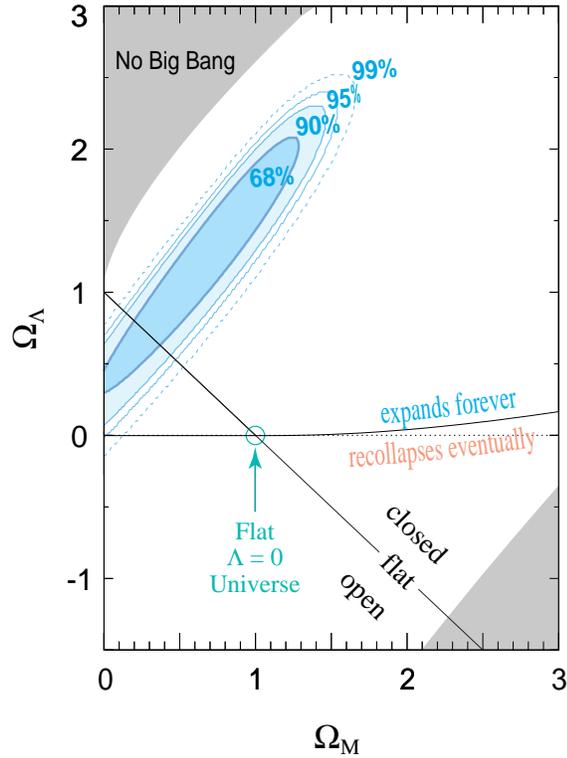}}
\caption{Cosmological versus matter densities in units of the flat universe
critical density \cite{ref:q}.}
\label{fig:14}
\end{center}
\end{figure}

Figure 14 shows the results in the matter-density/cosmological-constant
plane.  It is very far from being consistent with what we all had been
assuming for many years:  a matter density equal to one and a
cosmological constant of zero.  Personally, I would like to see this
analysis done in a slightly different fashion.  Since the universe
obviously has some matter in it, the matter-density should be
constrained to be greater than zero.  This constraint will rotate the
error ellipse and shift it somewhat.

The real question about the analysis has to do with the assumption that
Type 1a supernova are really standard candles, \ie, do supernovae that
exploded six or seven billion years ago ($z = 1$) give the same light
output as those that explode today?  Stars formed recently have more
heavy elements in them than did stars formed many billions of years ago,
and I don't know of any studies that look at what effect if any such a
systematic difference in composition might have on light output.  The
collaboration has clearly gone through the detailed analysis of
systematic errors, but I am not sure that I have high confidence in the
systematic error quoted of 0.05 on the mass density.  A systematic shift
of one-quarter to one-half magnitude at $z = 1$ would move the
cosmological constant to zero.  The data are superb, but the conclusions
may not be.  The collaboration will be collecting much more data and
have proposed a dedicated satellite to extend the data out to a $z$ of two
to three.  If I had the money, I would give it to them---this is really
important.

\section{Theory}

I want to talk about theory more philosophically than technically
because I think it important that experimenters understand theory better
so that they do not become mere technicians for those espousing the
latest theoretical fad.  Historically, advances in theory have
synthesized data, accommodated previous theory as a special case, and
simplified our view of the world.  We have a bias toward elegance (in
the eye of the beholder, of course) in the choice of theories and that
has served us well in the past.  The more mathematically and
conceptually elegant theories are the ones that tended to survive.

All theories are born under-constrained in that there are always
alternatives that compete in the contest with experiment, with the
losers consigned to the dustbin.  The sociologists of science would say
that our theories are ``socially constructed."  That is certainly true
initially but our theoretical models are continually tested and the
``social constructs" that don't pass are discarded; though sometimes it
takes a long time.  Aristotle and Democritus had different views on
whether matter was continuous or atomic in nature.  Although there was
no evidence to decide between these two views, Aristotle won, and for
2000 years it was believed that matter could be continuously subdivided
and that there were no such things as atoms.  We all believe that we are
much more critical these days in examining assumptions.

The Standard Model when born in the 1970s was thought to be good with
only one Higgs boson all the way up to the grand unification scale at
10$^{15}$ GeV.  Since that time it has accumulated problems.  It is hard to
suppress strong CP violation.  There is not enough CP violation to
reproduce the baryon asymmetry of the universe.  Longitudinal $W$
scattering is a problem if the Higgs mass is too high.  There is no way
to generate neutrino masses.  It has 18 constants without a lepton
sector CKM matrix and seven more with it.  While the standard model has
withstood all experimental tests, we know that it is only a low-energy
(a few hundred GeV) approximation to a better model.

The most popular candidate to be the successor to the plain vanilla
standard model is supersymmetry.  It was introduced to stabilize the
Higgs mass which is quadratically divergent in the standard model and
only logarithmically divergent in the supersymmetric variants of the
standard model.  Supersymmetry does reduce to the standard model at
``low" energies, but it also introduces 80 real and 44 complex constants.
The theorists who are fans of supersymmetry are groping for variants
that reduce these 124 new constants to a handful.  If the
supersymmetric successor to the standard model cannot reduce the total
number of constants, it would seem to me to be a step backwards rather
than an advance.

To the experimenters I would say that supersymmetry is a pure ``social
construct" with no supporting evidence despite many years of effort.  It
is okay to continue looking for supersymmetry as long as it doesn't
seriously interfere with real work (top, Higgs, neutrinos, \etc).

String/brane theory is in a very different situation.  It represents an
attempt to bring together gravity and quantum mechanics, a problem worth
serious effort.  It has too many dimensions for those of us living in a
four-dimensional world, but these are early days and perhaps they'll go
away in the appropriate limit.  There are several alternates that, on
further examination, seem to turn into each other through duality
transformations.  Hidden in the center of all of these alternates are
likely to be some kind of phase transitions that may lead to
experimental signatures like those found for inflation.  String/brane
theory may even give the necessary constraints that supersymmetry needs
to reduce the number of constants to a believable level.  It is still
too early to say, but it may be much more than metaphysics.

\section{Concluding Observations}

\begin{itemize}

\item[a)]
Experimenters (and phenomenologists) need to be more concerned about
systematic errors and the tails on error-distribution functions.

\item[b)]
Experimenters should learn more theory.

\item[c)]
All theorists should have a
required course in statistics before receiving their Ph.D.

\item[d)]
We all hope
for new things from LEP-II and the Tevatron, although the chances seem
small.  The new runs only increase LEP-II's mass reach by about 8 GeV
and the Tevatron's by 50 GeV.

\item[e)]
We have big hopes for the $B$-Factories.
The standard model's CP violation is not enough and new directions may
become clear from the factories.  The first results should be in next
summer.

\item[f)]
Neutrino physics is in ferment.  More Super-K data, SNO,
Borexino, KamLAND, Minos, K2K, should help to make things clear but it
will take four to five years.

\item[g)]
Some redundancy in neutrino experiments
is useful; too much is wasteful.

\item[h)]
I would love to see a low-energy
muon-neutrino disappearance and reappearance experiment.  Can it be done
with Minos, or an appropriately designed CERN-Gran Sasso experiment?

\item[i)]
LHC starts up in 2005 and we all hope to find out what is beyond our
standard model.  The experiments are huge and the sociology will be
complex.  Beware of too many boards and committees.

\item[j)]
An $e^+e^-$ collider of
0.5 to 1 TeV is a necessary companion to the LHC.  It will only come to
be if we all get behind it and push it as an international and regional
program.

\item[k)]
Muon colliders, VLHC's and exotic accelerator technologies are
machines for 2020 or beyond.  Muon colliders need R\&D work to
demonstrate that damping can work and then we'll still face formidable
problems.  VLHC's are a fantasy now.  In addition to cost breakthroughs
they need some serious accelerator physics studies (for example, already
at LHC some power supplies now need tolerances of one part in a
million).  Exotic techniques such as plasmas and lasers are still in
their infancy.  They have achieved accelerating gradients of 1 GeV per
meter, but only over a distance of a millimeter.

\item[l)]
Muon storage rings as
neutrino sources are interesting and much simpler than are muon
colliders.  A study of how to use polarization, mixed electron and muon
neutrino beams and their antiparticles is needed.  The machine and the
experiments interact strongly.

\item[m)]
Non-accelerator experiments in space and
on the ground will be of increasing importance.  The high-energy physics
community should not be too parochial.

\item[n)]
The string theorists are doing
great things.  I hope they justify or eliminate supersymmetry and think
up an experimental test.

\item[o)]
There is an exciting future:  the work will be
difficult, expensive and rewarding.  The young generation, with support
from governments, can and will do it.  This is not  ``The End of Science."

\end{itemize}


\begin{thebibliography}{99}

\bibitem{ref:a}
T. D. Lee and C. N. Yang, Phys. Rev. {\bf 104}, 254 (1956)

\bibitem{ref:b}
Bruno Pontecorvo, Inverse Beta Process, B. Pontecorvo, PD-205, Chalk River,
Ontario, Canada, November 1946.

\bibitem{ref:c}
Y. Suzuki, this conference.

\bibitem{ref:d}
L. DiLella, this conference.

\bibitem{ref:e}
LSND,  Phys.  Rev.  {\bf C54}, 2685 (1996).

\bibitem{ref:f}
T. Mann, this conference.

\bibitem{ref:g}
Fogli \etal,  hep-ph/9808205 and Phys.  Rev.   {\bf D59}, 033001 (1999).

\bibitem{ref:h}
Barger, \etal, Phys. Rev. Lett.  {\bf 82}, 2640 (1999).

\bibitem{ref:i}
Gonzales-Garcia, Phys. Rev. Lett.  {\bf 82}, 3202 (1999).

\bibitem{ref:j}
Barger, Geer and Whisnont ,  hep-ph 9906487.

\bibitem{ref:k}
P. Tsai, SLAC .


\bibitem{ref:l}
 J. Lykken, this conference.

\bibitem{ref:m}
G. A. Voss, this conference.

\bibitem{than}
T. Han, ``Mini-Workshop on Physics with the First Muon
Collider'', Fermilab, May 22, 1998; 
{\tt http://pheno.physics.wisc.edu/$\sim$than/files/fmc$\underline{\quad}$sm.ps}

\bibitem{ref:n}
Ankenbrandt \etal,  Phys Rev ST-Accel.  Beams 2 (1999), 081001.

\bibitem{ref:o}
C. Contaldi, \etal,  Phys.  Rev.  Lett.  {\bf 82} , 2034 (1999).

\bibitem{cobe}
From the COBE website at Lawrence Berkeley National Laboratory,
{\tt www.lbnl.gov}.

\bibitem{ref:q}
S. Perlmutter, this conference.


\bibitem{ref:p}
From M. Turner and J. Tyson,  Rev. Mod. Phys. {\bf 71}, S145 (1999).

\end{thebibliography}
\end{document}